\documentclass[aip,reprint,jcp]{revtex4-1}
\usepackage{graphicx}
\usepackage{amsmath}
\usepackage{amssymb}
\usepackage{chemarr}
\usepackage[english]{babel}
\usepackage{color}
\usepackage{float}
\usepackage{array}
\usepackage{booktabs}
\usepackage{multirow}
\usepackage{threeparttable}

\begin{document}

\title{Approximate calculation of the binding energy between 17$\beta$-estradiol and human estrogen receptor alpha}

\author{Ricardo Ugarte}

\email{rugarte@uach.cl}

\affiliation{Instituto de Ciencias Quimicas, Facultad de Ciencias, Universidad Austral de Chile. Independencia 641, Valdivia, Chile.}

\begin{abstract}

Estrogen receptors (ERs) are a group of proteins activated by 17$\beta$-estradiol. The endocrine-disrupting chemicals (EDCs) mimic estrogen action by bind directly to the ligand binding domain of ER. From this perspective, ER represent a good model for identifying and assessing the health risk of potential EDCs. This ability is best reflected by the ligand-ER binding energy. Multilayer fragment molecular orbital (MFMO) calculations were performed which allowed us to obtain the binding energy using a calculation scheme that considers the molecular interactions that occur on the following model systems: the bound and free receptor, 17$\beta$-estradiol and a water cluster. The bound and free receptor and 17$\beta$-estradiol were surrounded by a water shell containing the same number of molecules as the water cluster. The structures required for MFMO calculations were obtained from molecular dynamics simulations and cluster analysis. Attractive dispersion interactions were observed between 17$\beta$-estradiol and the binding site hydrophobic residues. In addition, strong electrostatic interactions were found between 17$\beta$-estradiol and the following charged/polarized residues: Glu 353, His 524 and Arg 394. The FMO2-RHF/STO-3G:MP2/6-31G(d) weighted binding energy was of -67.2 kcal/mol. We hope that the model developed in this study can be useful for identifying and assessing the health risk of potential EDCs.

\end{abstract}

\maketitle

\section{Introduction} 

The chemical industry is introducing around 700 synthetic compounds on the market every year \cite{1}. These chemicals come on top of the 85000 substances listed in the EPA's chemical inventory. These compounds have a very wide range of applications and humans will come in contact to most of them through various routes. They may take a long time to degrade into harmless products. Some may not break down and persist in the environment. A large number of synthetic chemicals have been shown to damage wildlife populations, and pose large-scale hazards to human health \cite{2}. Despite their negative effects, humanity is increasingly dependent on synthetic chemicals. According to the UN, output will grow seven times faster than the global population between 1990 and 2030 \cite{3}. This “chemical explosion” is perhaps one of the most formidable challenges confronting mankind today \cite{4,5}.

Nuclear receptors (NRs) are evolutionary conserved intracellular proteins responsible for transmitting external signals to the cell nucleus. Most of these proteins act as ligand-inducible transcription factors and respond to endogenous (endobiotics) and exogenous (xenobiotics) chemicals in order to regulate gene expression. NRs affect a variety of biological functions, such as reproductive development or detoxification of foreign substances and fatty acid metabolism. NRs mediate chemical communication between different organs via the endocrine system, but also the interaction between organisms and their environment. They act as xenosensors and endocrine regulators, which connect and integrate endogenous hormone-regulated functions with external dietary and/or environmental stimuli \cite{6}.

\begin{figure}[th]
\includegraphics[width=0.65\columnwidth]{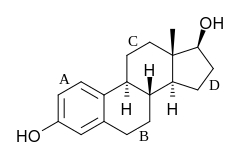}
\caption{\small Chemical structure of 17$\beta$-estradiol ($C_{18}H_{24}O_{2}$)}
\label{figure1}
\end{figure}

The endocrine system is sensitive to stimulations by low concentrations of hormones. Chemicals acting as endocrine-disrupting chemicals (EDCs) \cite{7}, either natural or synthetic, alters the hormonal and homeostatic systems that enable the organism to communicate with and respond to its environment. EDCs affect the endocrine system by mimicking natural hormones, antagonizing their action or modifying their synthesis, metabolism and transport. A large number of industrial chemicals have polycyclic aromatic structures which confer them the ability to bind NRs involved in steroid hormone metabolism. These EDCs include synthetic chemicals used as industrial solvents/lubricants and their byproducts (polychlorinated biphenyls, polybrominated biphenyls, dioxins), plastics (bisphenol A), plasticizers (phthalates), pesticides (methoxychlor, dichlorodiphenyltrichloroethane or DDT), fungicides (vinclozolin), and pharmaceutical agents (diethylstilbestrol). Natural chemicals found in human and animal food chains (e.g., phytoestrogens, including genistein and coumestrol) can also act as endocrine disruptors \cite{8}. 

A large range of xenobiotics have been found to bind and activate estrogen receptors (ERs), a group of proteins activated by the sex steroid hormone 17$\beta$-estradiol (E2) (Figure 1). Two subtypes of ER exist: ER$\alpha$ and ER$\beta$, which are members of the nuclear receptors superfamily. Both ER subtypes possess a modular organization that is characteristic of the NRs; five functional domains from the N- to C-termini, designated A/B, C (DNA-binding domain, DBD), D, E (ligand-binding domain, LBD), and F \cite{9,10}. Numerous crystal structures have been determined for the LBDs of both subtypes, and these have given a detailed insight into the structure and alterations during the ligand binding. The structure of ER LBD reveals a conserved core of twelve $\alpha$-helices and a short two-stranded antiparallel $\beta$-sheet arranged into a three-layered sandwich fold; this arrangement generates a mostly hydrophobic cavity in the lower part of the domain which can accommodate the ligand \cite{6} (Figure 2). Since different classes of compounds might bind to ER LBD and elicit hormone-like effects in humans, ER represent a good model system for identifying and assessing the health risk of potential EDCs. This ability is best reflected by the ligand-ER binding energy. A number of experimental and theoretical studies have been performed to investigate the ligand-ER interaction \cite{11}$^{-}$\cite{34}, and since 1997 about 100 crystal structures of ER LBD with several ligands have been solved and deposited in the Research Collaboratory for Structural Bioinformatics (RCSB) Protein Data Bank (PDB). On the basis of the above information, the mode of binding between ERs and their ligands has been determined. The specific recognition between ER and its ligand mainly depends on hydrogen bonds and hydrophobic contacts \cite{35,36}.

Most of the theoretical studies which use the structures deposited in RCSB PDB are carried out using molecular dynamics (MD) simulations. These simulations are based on empirical force fields that may not be accurate enough to predict ligand-ER binding energies. Accuracy requirements could be provided by ab initio quantum mechanical calculations, but these can be very computationally expensive and time consuming. The hybrid QM/MM (quantum mechanics/molecular mechanics) is a method that combines the precision of quantum mechanics and the speed of empirical force fields. In this approach, part of the system that includes the chemically relevant region is treated quantum mechanically (QM) while the remainder, often referred to as the environment, is treated at the classical level using empirical or molecular mechanics (MM) force fields. This multiscale approach reduces the computational cost significantly as compared to a QM treatment of the entire system and makes simulations possible \cite{38,39}. An efficient alternative to either the full ab initio QM, MM or QM/MM, lies in the fragment-based methods, which form an actively developed field of research \cite{40}. Fragment Molecular Orbital (FMO) method \cite{41,42} is one such method that has been used for efficient and accurate QM calculations in very large molecular systems \cite{21,43}. FMO involve fragmentation of the chemical system, and from ab initio or density functional quantum-mechanical calculations of each fragments (monomers) and their dimers (and trimers if greater accuracy is required) one can construct the total properties. The distinctive feature of FMO is the inclusion of the electrostatic field from the whole system in each individual fragment calculation, and in using the systematic many-body expansion. The FMO method is suited to various analyses, as it provides information on fragments and their interactions that are naturally built into the method.

In the present article, we report a study on an approximate calculation of the binding energy between 17$\beta$-estradiol and LBD of human estrogen receptor alpha in aqueous medium. Briefly, the main steps of the calculation are as follows: 1) Search for representative structures of the conformational space around the crystallographic state of E2-ER$\alpha$ LBD by means of MD simulations and cluster analysis; 2) Geometry optimization of the representative structures using QM/MM approach; 3) Single point FMO calculation on the optimized structures in order to obtain the inter-fragment interaction energies. In general, the same steps were applied to other related model systems (vide infra).

\begin{figure}[t]
\includegraphics[width=0.40\columnwidth]{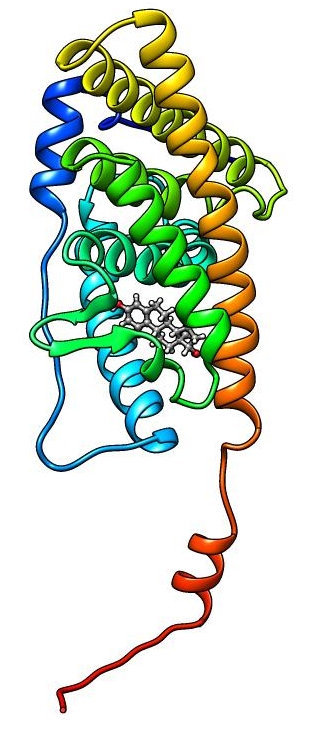}
\caption{\small Model of ER$\alpha$ LBD (ribbon) complexed with EST (ball and stick) \cite{37}. The model based on the RSCB PDB crystal structure (PDB code 1A52) includes 258 amino acid residues.}
\label{figure2}
\end{figure}

\section{Methods}

\subsection{Model Building}

Crystal structure of the ER$\alpha$ LBD in complex with E2 (PDB code 1A52) were downloaded from the RSCB PDB \cite{44}. The model was constructed from chain A of the homodimer. Atomic coordinates for missing amino acid residues (297-305, 545-554), missing heavy atoms and hydrogen atoms, were reconstructed with the LEaP module of AmberTools 15 package \cite{45}. AMBER FF14SB force field was selected for the proteins and general AMBER force field (GAFF) parameters \cite{46} were employed for E2. In order to parameterize E2, electrostatic potential was calculated by Gaussian 09 program at the HF/6-31G(d) level of theory \cite{47}. Partial charges were fitted by RESP method of the Antechamber module of AmberTools 15 \cite{48}. Arg, Lys, Asp, Glu residues were modeled as charged species, all tyrosines as neutral, and histidine residues were modeled according to information obtained from another source \cite{22}. Three of the 13 residues of histidine, were protonated in order to preserve the electroneutrality of the system. The N- and C-terminus residues were protonated and deprotonated, respectively. The model was solvated with TIP3P water in a pre-equilibrated box measuring 111 x 100 x 62 \r{A}$^{3}$. The E2-ER-W system contains 258 amino acid residues (4190 atoms), E2 and 19979 water molecules (W). The total number of atoms in the system is 64171. E2-ER-W was subjected to three successive steps of minimization using the SANDER module of the AmberTools 15. First, 1000 steps of steepest descent followed of 1500 steps of conjugate gradient, allowing only H atoms and water to move while holding the rest of the system fixed. Next, the same minimization algorithm is used as the previous one, but allowing only E2-ER-W to move. Finally, the whole system was minimized without any restraints for 2000 steps of steepest descent followed by 1000 steps of conjugate gradient.
 
\begin{figure*}
\includegraphics[width=1.70\columnwidth]{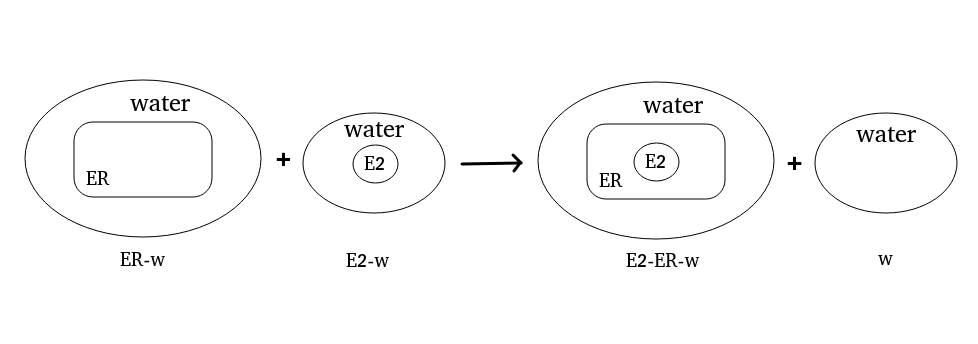}
\caption{\small FMO2 calculation model: $ \Delta E_{Binding} = \Big[\Delta E^{E2-ER-w}_{Interaction} + \Delta E^{w}_{Interaction}\Big] - \Big[\Delta E^{ER-w}_{Interaction} + \Delta E^{E2-w}_{Interaction}\Big] $}
\label{figure3}
\end{figure*}

\subsection{Molecular Dynamics Simulations}

All simulations were carried out with the SANDER module of the AmberTools 15 with periodic boundary conditions, using Particle Mesh Ewald method \cite{49} to treat long-range electrostatics interactions with a non-bonded cutoff of 12 \r{A}. All bonds involving hydrogen atoms were restrained using the SHAKE \cite{50} algorithm. Temperature regulation was done using a Langevin thermostat with collision frequency of 1 ps$^{-1}$. The Berendsen barostat was used for constant pressure simulation at 1 atm, with a relaxation time of 1 ps. The time step was 1 or 2 fs. The final energy-minimized system (E2-ER-W) was then submitted to the following protocol: 
\\
\\
Scheme 1: [0$\rightarrow$310 K: 100 ps  1 fs  NVT]$\longrightarrow$[310 K: 500 ps 2 fs  NPT]$\longrightarrow$[310$\rightarrow$5 K: 100 ps  1 fs  NVT]
\\

From the restart file of the last simulation in Scheme 1, we conducted an extensive set of molecular dynamics simulations to explore the conformational space in the vicinity of the crystallographic structure. To circumvent the limited conformational sampling ability of MD simulations at 310 K, we used multiple-trajectory short-time simulations \cite{51}. By combining the sampling ability of the multiple trajectories, we expect to sample more conformational space than single trajectory of the same length. The aforementioned restart file was used as seed for 30 short-time simulations that obey the protocol established in Scheme 2:
\\
\\
Scheme 2: [5$\rightarrow$150 K: 30 ps  1 fs  NVT]$\longrightarrow$[150$\rightarrow$310 K: 70 ps  1 fs   NPT]$\longrightarrow$[310 K: 500 ps  1 fs  NVE]
\\

The initial velocities (Scheme 2) were assigned randomly from a Maxwell-Boltzmann distribution at 5 K. The trajectories start with the same structure and differ only in the initial velocity assignment. At the end of the equilibration, from $\sim$ 40 ps NPT ensemble, the average temperature of the final 30 ps was 310 K, and the average density was 1.0 g/ml. All production runs of 0.5 ns were performed in an NVE ensemble at 310 K.

\subsection{Cluster Analysis}

Cluster analysis methods have been developed for analyzing simulation trajectories of biomacromolecules and used for the analysis of their conformational behavior in solution \cite{52,53}. These methods group together similar conformers from molecular simulations. A clustering approach based on the C$\alpha$-RMSD (root mean square deviation) was applied to the snapshots of the MD simulations. We selected the alpha carbon atoms because they describe the backbone conformations. The C$\alpha$-RMSD was calculated after rigid body alignment of C$\alpha$ atoms of each frame of the trajectory with respect to C$\alpha$ atoms of the average structure of the protein. Prior to clustering, the individual E2-ER-W trajectories from the 30 short-time simulations were combined into a single file and the water molecules removed. In our analysis, 15000 snapshots were grouped into five clusters. Each cluster is described by a centroid structure, which in itself is not physically significant as it is effectively a mathematical construct based on the members of a cluster. However, the actual structure closest to the centroid (rmsd) is significant and representative of each cluster. Therefore, each of these structures corresponds to snapshots of the MD trajectory. Thus, five representative structures (E2-ER) of each cluster, and therefore of the conformers population, were obtained \cite{54}. Finally, from E2-ER we return to the corresponding E2-ER-W representative structures.
	
\subsection{QM/MM Optimization}

All E2-ER-W systems representative of the population were subjected to geometry optimization using Gaussian 09 at the MM/Amber level of theory. Then, to facilitate the QM/MM calculations, the water molecules beyond 10 \r{A} of the protein surface were deleted using VMD program \cite{55}. As a consequence, new model systems (E2-ER-w) with a water layer of 10 \r{A} around of receptor surface were generated. ONIOM, \cite{38} a hybrid QM/MM method implemented in Gaussian 09, with electronic embedding was used for the geometry optimization of E2-ER-w model systems. Electronic embedding procedure best describes the electrostatic interaction between the QM (E2) and MM (ER-w) regions, because it includes the partial charges of the MM region into the quantum mechanical Hamiltonian. In this way, the wave function of the QM region can be polarized. In the present study we used a two-layer ONIOM (B3LYP/6-31+G(d):AMBER) scheme: E2 (B3LYP/6-31+G(d)); ER-w(AMBER).

\subsection{Related Model Systems}

In order to calculate the binding energy we also require ER-w, E2-w and water (w) model systems (Figure 3). ER-w and E2-w are derived, respectively, of MD simulations of ER-W and E2-W systems. Practically the same protocol described in the methods section was required to obtain five representative ER-w structures and two representative E2-w structures. In addition, by removing 17$\beta$-estradiol from E2-ER-w and E2-w followed by optimizing the geometries of the resulting systems at the appropriate level of theory, five new ER-w and two w model systems were obtained.

\subsection{FMO Calculations}

Fragment-based approaches allude to the chemical idea of parts of the system retaining their identity to a large extent (e.g., functional groups and residues). FMO method not only reduces the computational cost, but it also provides a wealth of information on the properties of fragments and their interactions. In this study, the calculation of the binding energy is based on obtaining the following interaction energies between pair of fragments: E2-amino acid residue (est-aa), E2-water (est-wat), amino acid residue-amino acid residue (aa-aa), amino acid residue-water (aa-wat) and water-water (wat-wat). These pair interaction energies (PIEs) will be computed in the following model systems: ER-w, E2-w, E2-ER-w and w (Figure 3).

The energy expression in the two-body FMO expansion (FMO2) is \cite{41}:

\begin{equation}
E = \sum_{I}^{N}E_{I} + \sum_{I>J}^{N}(E_{IJ}-E_{I}-E_{J})
\label{1}
\end{equation}

The total energy $E$ of the system is written as the sum of the monomer energies $E_{I}$, and the pair interaction energies $(E_{IJ} - E_{I} - E_{J}  = \Delta E_{IJ})$, where $E_{IJ}$ is the energy of the dimer made of two fragments $I$ and $J$. The order of the fragments in the FMO input file is very important: water $\rightarrow$ ER amino acid residues $\rightarrow$ 17$\beta$-estradiol. Let A = number of water fragments, B = number of {water + amino acid residue} fragments and C = total number of fragments (water + amino acid residue + E2). According to the above schema and the FMO2 calculation model (Figure 3):

\begin{equation}
\Delta E_{Binding} = [E^{E2-ER-w} + E^{w}] - [E^{ER-w} + E^{E2-w}]
\label{2}
\end{equation}

\begin{equation}
E^{E2-ER-w} = \sum_{I=1}^{N=C}E_{I} + \sum_{J=1 \ I>J}^{N=C}\Delta E_{IJ}
\label{3}
\end{equation}

\begin{equation}
E^{w} = \sum_{I=1}^{N=A}E_{I} + \sum_{J=1 \ I>J}^{N=A}\Delta E_{IJ}
\label{4}
\end{equation}

\begin{equation}
E^{ER-w} = \sum_{I=1}^{N=B}E_{I} + \sum_{J=1 \ I>J}^{N=B}\Delta E_{IJ}
\label{5}
\end{equation}     

\begin{multline}
E^{E2-w} = \sum_{I=1}^{N=A}E_{I} + E_{A+1} + \sum_{J=1 \ I>J}^{N=A}\Delta E_{IJ}\\ + \sum_{J=1}^{N=A}\Delta E_{(A+1)J}
\label{6}
\end{multline}

We assume that the monomer energies are practically independent of the model systems and since these terms are subtracted from each other, then:

\begin{multline}
\Delta E_{Binding} = \sum_{J=1 \ I>J}^{N=C}\Delta E_{IJ} + \sum_{J=1 \ I>J}^{N=A}\Delta E_{IJ}\\ - \sum_{J=1 \ I>J}^{N=B}\Delta E_{IJ} - \sum_{J=1 \ I>J}^{N=A}\Delta E_{IJ} - \sum_{J=1}^{N=A}\Delta E_{(A+1)J}
\label{7}
\end{multline}

The right-side terms of equation (7) represent sums of pair interaction energies (PIEs). For the purpose of simplifying notation:

\begin{multline}
\Delta E_{Binding} = \Big[\Delta E^{E2-ER-w}_{Interaction} + \Delta E^{w}_{Interaction}\Big]\\ - \Big[\Delta E^{ER-w}_{Interaction} + \Delta E^{E2-w}_{Interaction}\Big]
\label{8}
\end{multline}

From the FMO output file we obtain the values of the terms on the right-side of the previous equation.  The AFO (adaptive frozen orbitals) scheme was used throughout for fragmentation across peptide bonds, with the default settings for bond definitions. The fragmentation of the model system was as follows: The first two amino acid residues and each remaining amino acid residue of ER, 17$\beta$-estradiol molecule, and the water molecule were treated as a single fragment. Table I shows the number of fragments in the different model systems.

\begin{table}
 \begin{threeparttable}
\caption{Number of fragments in the model systems}
 \begin{tabular}{lllll}
 \toprule[1pt] 
    Fragments \ \ & ER-w \ & E2-w & \ E2-ER-w & \ \ w \\
     \midrule[1pt]
 17$\beta$-estradiol \ \ &  & \ \ \ \ 1 & \ \ \ \ \ \ \ 1 & \\
   Amino acid residue \ \  & \ 257 &  & \ \ \ \ 257 & \\
          Water \ \ & 5153 & 5153 & \ \ \ 5153 & 5153 \\
 \midrule[1pt]
          Total \ \ & 5410 & 5154 & \ \ \ 5411 & 5153 \\ 
 \bottomrule[1pt]
 \end{tabular}
     \begin{tablenotes}
      \small
      \item
    \end{tablenotes}
\label{Table I}
  \end{threeparttable}
\end{table}

For the binding energy ($\Delta E_{Binding}$) calculation the part of the system that is of particular interest corresponds to the ligand and the binding site. In FMO, one can address this by using multilayer FMO (MFMO), when several fragments are assigned to a higher layer. Wavefunctions and basis sets can be defined separately for each layer. In this work we used the two-layer two-body FMO method: FMO2-RHF/STO-3G:MP2/3-21G and FMO2-RHF/STO-3G:MP2/6-31G(d) level of theory. For example, the latter means the two-layer two-body FMO method with layer 1 (environment) described by RHF and the STO-3G basis set, and layer 2 (E2 and binding site) described by MP2 and the 6-31G(d) basis set \cite{56}.

The binding site consists of all residues that have at least one atom within 3.5 \r{A} from any 17$\beta$-estradiol atom in E2-ER-w. This generally gives a good representation of the important residues in the binding pocket of a protein. The amino acids residues that form the binding site of E2-ER-w and ER-w are: Leu 346, Leu 349, Ala 350, Glu 353, Leu 384, Leu 387, Met 388, Leu 391, Arg 394, Phe 404, Met 421, Ile 424, Leu 428, Gly 521, His 524, Leu 525, Met 528. The binding site was constructed by using the ArgusLab software \cite{57}.

\section{Results and Discussion}

In cluster analysis, 15000 snapshots from MD simulations were processed and grouped into five (E2-ER-w, ER-w) and two (E2-w) clusters. Thus, representative structures (RS) of each cluster, and therefore of the conformers population, were obtained (Table II). The population of each cluster is important for the calculation of a weighted binding energy.

\subsection{Binding energy using ER-w structures from MD simulations of ER-W system}

Table II shows the cluster population of each system and the symbol assigned to their representative structures. Because E2 is a relatively rigid molecule, its population of conformers could be described by only two RS (X, Y). In order to calculate the binding energy, all possible combinations between RS were made. For example, $A\omega_{X}1X$ symbolizes the following calculation scheme (Figure 3): $1 + X \rightarrow A + \omega_{X}$. In this notation $\omega_{X}$ is the water representative structure obtained from X.

\begin{table*}[!htbp]
  \begin{threeparttable}
\caption{Cluster Analysis of MD trajectories in the different model systems (MS).}
\begin{tabular}{ccccccccccccc}
  \toprule[1pt]
  & \multicolumn{5}{c}{{E2-ER-w}}
  & \multicolumn{5}{c}{ \ {ER-w}}
  & \multicolumn{2}{c}{ \ \ {E2-w}} \\
  \midrule[1pt]
  CP$^{(a)}$ & 10215 & 3076 & 282 & 1214 & 213 & \ \ \ 7843 & 3577 & 2066 & 249 & 1265 & \ \ \ 5873 & 9127 \\
  RS$^{(b)}$ & \ A & B & C & D & E & \ \ \ 1 & 2 & 3 & 4 & 5 & \ \ \ X & Y \\
  \bottomrule[1pt]
\end{tabular}
\begin{tablenotes}
      \small
      \item $^{(a)}$Cluster population. $^{(b)}$Representative structure of the respective cluster.
\end{tablenotes}
\label{Table II}
  \end{threeparttable}
\end{table*}

Table III shows the binding energies calculated at the FMO2-RHF/STO-3G:MP2/3-21G level of theory. As we can see there is a large dispersion in the binding energy obtained with the proposed calculation scheme; the extreme values are the result of the remarkable difference of the interaction energies, $ \Delta E^{E2-ER-w}_{Interaction} - \Delta E^{ER-w}_{Interaction} $, between some of the RS (data not shown). The basis of this dispersion is structural and is corroborated by the protein backbone RMSD between pairs of superimposed representative structures: E2-ER-w//ER-w. The RMSD across all 258 pairs of amino acids on the twenty five combinations (A//1, A//2... E//5) of the RS was measured. By averaging all these measurements, the calculated RMSD mean value is 5.0 $\pm$ 1.6 \r{A}, a very high value as a result of high conformational variability of the terminal ends of the protein \cite{58}. This flexibility of the terminal ends has a strong impact on the aa-aa, aa-wat and wat-wat interactions.

Although the calculation scheme combining the representative structures obtained through the MD simulations (except for $\omega_{X}$ or $\omega_{Y}$ which derives directly from X or Y) seems reasonable and unbiased, it fails to calculate the binding energy.

\begin{table*}[!htbp]
  \begin{threeparttable}
\caption{FMO2-RHF/STO-3G:MP2/3-21G binding energy$^{(a)}$ (BE) of the different combinations between the representative structures.}
\begin{tabular}{cccccccccc}
  \toprule[1pt]
Model & BE \ \ & Model & BE \ \ & Model & BE \ \ & Model & BE \ \ & Model & BE \\
  \midrule[1pt]
$A\omega_{X}1X$ & 215.5 \ \ & $B\omega_{X}1X$ & 179.7 \ \ & $C\omega_{X}1X$ & 2.1 \ \ & $D\omega_{X}1X$ & 381.0 \ \ & $E\omega_{X}1X$ & -46.1 \\
$A\omega_{Y}1Y$ & 197.3 \ \ & $B\omega_{Y}1Y$ & 161.5 \ \ & $C\omega_{Y}1Y$ & -16.1 \ \ & $D\omega_{Y}1Y$ & 362.8 \ \ & $E\omega_{Y}1Y$ & -64.3 \\
$A\omega_{X}2X$ & 126.1 \ \ & $B\omega_{X}2X$ & 90.3 \ \ & $C\omega_{X}2X$ & -87.3 \ \ & $D\omega_{X}2X$ & 291.6 \ \ & $E\omega_{X}2X$ & -135.5 \\
$A\omega_{Y}2Y$ & 107.9 \ \ & $B\omega_{Y}2Y$ & 72.1 \ \ & $C\omega_{Y}2Y$ & -105.5 \ \ & $D\omega_{Y}2Y$ & 273.4 \ \ & $E\omega_{Y}2Y$ & -153.7 \\
$A\omega_{X}3X$ & -95.3 \ \ & $B\omega_{X}3X$ & -131.1 \ \ & $C\omega_{X}3X$ & -308.7 \ \ & $D\omega_{X}3X$ & 70.2 \ \ & $E\omega_{X}3X$ & -356.9 \\
$A\omega_{Y}3Y$ & -113.5 \ \ & $B\omega_{Y}3Y$ & -149.3 \ \ & $C\omega_{Y}3Y$ & -326.9 \ \ & $D\omega_{Y}3Y$ & 52.0 \ \ & $E\omega_{Y}3Y$ & -375.1 \\
$A\omega_{X}4X$ & 59.1 \ \ & $B\omega_{X}4X$ & 23.3 \ \ & $C\omega_{X}4X$ & -154.3 \ \ & $D\omega_{X}4X$ & 224.6 \ \ & $E\omega_{X}4X$ & -202.5 \\
$A\omega_{Y}4Y$ & 40.9 \ \ & $B\omega_{Y}4Y$ & 5.1 \ \ & $C\omega_{Y}4Y$ & -172.5 \ \ & $D\omega_{Y}4Y$ & 206.4 \ \ & $E\omega_{Y}4Y$ & -220.7 \\
$A\omega_{X}5X$ & -13.6 \ \ & $B\omega_{X}5X$ & -49.4 \ \ & $C\omega_{X}5X$ & -227.0 \ \ & $D\omega_{X}5X$ & 151.9 \ \ & $E\omega_{X}5X$ & -275.2 \\          
$A\omega_{Y}5Y$ & -31.8 \ \ & $B\omega_{Y}5Y$ & -67.6 \ \ & $C\omega_{Y}5Y$ & -245.2 \ \ & $D\omega_{Y}5Y$ & 133.7 \ \ & $E\omega_{Y}5Y$ & -293.4 \\
  \bottomrule[1pt]
\end{tabular}
\begin{tablenotes}
      \small
      \item $^{(a)}$All Energies in kcal/mol.
\end{tablenotes}
\label{Table III}
  \end{threeparttable}
\end{table*}

\subsection{Binding energies using ER-w structures from of E2-ER-w model system}

Table IV shows the binding energies calculated at the FMO2-RHF/STO-3G:MP2/3-21G and FMO2-RHF/STO-3G:MP2/6-31G(d) level of theory. The binding energy calculation schema is analogous to the above. For example, $A\omega_{X}aX$ symbolizes the following calculation scheme: $a + X \rightarrow A + \omega_{X}$. Here, "a" (without quotes) stands for the representative structure derived from A by prior elimination of 17$\beta$-estradiol.

All binding energy values are negative which, at least at this level of calculation, would indicate a certain stability of the system; furthermore, the dispersion in the values lie within a normal range. The RMSD mean value obtained from the RMSD measurements on the five ensembles (A//a... E//e) is 0.035 $\pm$ 0.0063 \r{A}, which is consistent with the dispersion observed in the binding energy values. 

The interaction energy and, therefore, the binding energy calculated by each basis set differ significantly. One of the reasons for this basis set dependence is due to no correction of the basis set superposition error (BSSE). This problem that manifests mainly on smaller basis sets, such as STO-3G and 3-21G, tends to overestimate the interaction energy \cite{59}.

\begin{table*}[!htbp]
  \begin{threeparttable}
\caption{FMO2-RHF/STO-3G:MP2/3-21G \& FMO2-RHF/STO-3G:MP2/6-31G(d) binding (BE) and interaction energy$^{(a)}$ ($\Delta E_{int}$).}
\begin{tabular}{ccccccccccc}
  \toprule[1pt]
       \multicolumn{11}{c}{FMO2-RHF/STO-3G:MP2/3-21G} \\
      \multicolumn{11}{c}{Weighted Binding Energy = -84.2} \\
     \multicolumn{11}{c}{Weighted $\Delta E_{int}$ = -92.3} \\
  \hline
  Model & $A\omega_{X}aX$ & $A\omega_{Y}aY$ & $B\omega_{X}bX$ & $B\omega_{Y}bY$ & $C\omega_{X}cX$ & $C\omega_{Y}cY$ & $D\omega_{X}dX$ & $D\omega_{Y}dY$ & $E\omega_{X}eX$ & $E\omega_{Y}eY$ \\
  BE & -71.7 & -89.9 & -73.7 & -91.9 & -80.5 & -98.7 & -79.7 & -97.9 & -85.3 & -103.5 \\
  \hline
  \ \ \ RS$^{(b)}$ & \multicolumn{2}{c}{A} & \multicolumn{2}{c}{B} & \multicolumn{2}{c}{C} & \multicolumn{2}{c}{D} & \multicolumn{2}{c}{E} \\
$\Delta E_{int}$ & \multicolumn{2}{c}{-93.0} & \multicolumn{2}{c}{-88.6} & \multicolumn{2}{c}{-105.7} & \multicolumn{2}{c}{-91.3} & \multicolumn{2}{c}{-101.8} \\ 
\midrule[1pt]
\\
\midrule[1pt] 
   \multicolumn{11}{c}{FMO2-RHF/STO-3G:MP2/6-31G(d)} \\
  \multicolumn{11}{c}{Weighted Binding Energy = -67.2} \\
  \multicolumn{11}{c}{Weighted $\Delta E_{int}$ = -73.6} \\
  \hline
  Model & $A\omega_{X}aX$ & $A\omega_{Y}aY$ & $B\omega_{X}bX$ & $B\omega_{Y}bY$ & $C\omega_{X}cX$ & $C\omega_{Y}cY$ & $D\omega_{X}dX$ & $D\omega_{Y}dY$ & $E\omega_{X}eX$ & $E\omega_{Y}eY$ \\
  BE & -54.6 & -72.8 & -57.4 & -75.6 & -66.9 & -85.1 & -62.1 & -80.3 & -64.0 & -82.2 \\
  \hline  
  RS & \multicolumn{2}{c}{A} & \multicolumn{2}{c}{B} & \multicolumn{2}{c}{C} & \multicolumn{2}{c}{D} & \multicolumn{2}{c}{E} \\
$\Delta E_{int}$ & \multicolumn{2}{c}{-74.3} & \multicolumn{2}{c}{-69.6} & \multicolumn{2}{c}{-85.4} & \multicolumn{2}{c}{-74.0} & \multicolumn{2}{c}{-79.2} \\ 
  \bottomrule[1pt]   
\end{tabular}
\begin{tablenotes}
      \small
      \item $^{(a)}$Sum of all PIEs between 17$\beta$-estradiol and each amino acid residue fragment in the ER binding site. All Energies in kcal/mol. $^{(b)}$Representative structure.
\end{tablenotes}
\label{Table IV}
  \end{threeparttable}
\end{table*}

In an FMO study of ER$\alpha$ LBD in complex with 17$\beta$-estradiol (PDB code 1ERE), the model system included 241 amino acid residues, one water molecule which directly mediates ER-E2 binding (where the hydrogen bonded water molecule was included in the receptor) and E2. The binding energy was estimated from: $\Delta E_{Binding} = E_{E2-ER} - (E_{ER} + E_{E2})$. The total energies (E) were considered in the calculation and the geometries of ER and E2 were fixed in those found in E2-ER model system. The reported binding energy was -37.65 kcal/mol at FMO2-RHF/STO-3G level of theory \cite{21}. In another FMO study with the same model system, FMO2 interaction energy between E2 and ER was calculated using HF and MP2 methods with several basis sets \cite{60}. The calculated interaction energy was -40.26 kcal/mol at FMO2-RHF/6-31G(d) and -123.73 kcal/mol at FMO2-MP2/6-31G(d) level of theory. The large interaction energy difference between the RHF and the MP2 methods is due to dispersion interaction, which can only be described by electron correlation methods. In general, charged and polarized amino acid residues interact either strongly or weakly with the ligand, while hydrophobic residues contribute to weak interactions. The sum of these weak dispersion interactions makes the difference between both methods.

\begin{figure}[t]
\includegraphics[width=1.05\columnwidth]{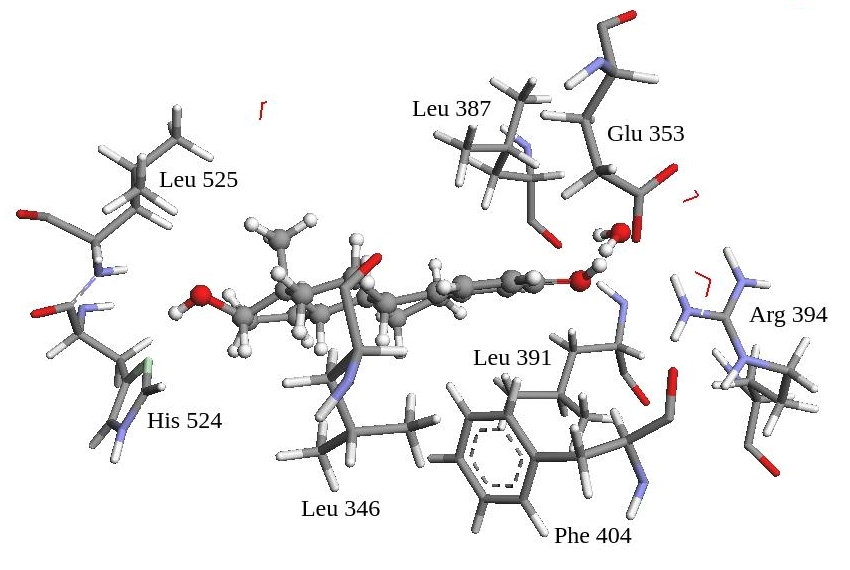}
\caption{17$\beta$-estradiol in the binding site of ER with important amino acid residues. Four water molecules are present, and one of them (cylinder) links E2 and Leu 387. The figure corresponds to the A representative structure and it was made with the ArgusLab software \cite{57}.}
\label{figure4}
\end{figure}

The interaction energies of E2 with each residue fragment of the ER binding site at the FMO2-RHF/STO-3G:MP2/3-21G and FMO2-RHF/STO-3G:MP2/6-31G(d) level of theory are shown in Figure 5. In all model systems, regardless of the calculation level, we observe that Glu 353 and His 524 are very stable structures. These residues present a strong electrostatic interaction with E2, and it is known that together with Arg 394, they form a hydrogen bond network with E2. A water molecule is similarly responsible for yet another stabilizing hydrogen bond between E2 and the ER (Figure 4) \cite{22,60}. Therefore, the interactions between E2 and these residues, together with the hydrogen bonds, play a key role in the E2-ER binding.

Many binding site hydrophobic residues are stabilized (attractive interaction) through dispersion interactions with E2. Therefore, the MP2 electron correlation is essential to characterize these interactions, whose function is possibly to accommodate the substrate at the hydrophobic binding pocket (Figure 4). Important hydrophobic residues at the binding site are: Leu 346, Leu 387, Leu 391, Phe 404 and Leu 525 \cite{61}. The behaviour of the Arg 394 PIE (Figures 5) with respect to the rest of the amino acid residues is remarkable. In the latter, the behaviour is quite conservative; whereas in Arg 394 it fluctuates from -12.4 to 5.0 kcal/mol and from -7.5 to 4.6 kcal/mol, at FMO2-RHF/STO-3G:MP2/3-21G and FMO2-RHF/STO-3G:MP2/6-31G(d) level of theory, respectively. The above suggests that perhaps Arg 394 could play a stabilizing-destabilizing role in the interaction between E2 and the ER binding site. According to this hypothesis, conformational changes in the receptor could influence the behavior of Arg 394, which would possibly determine both the entry and exit of E2 from the binding site.
 
\begin{figure*}[t]
\includegraphics[width=2.15\columnwidth]{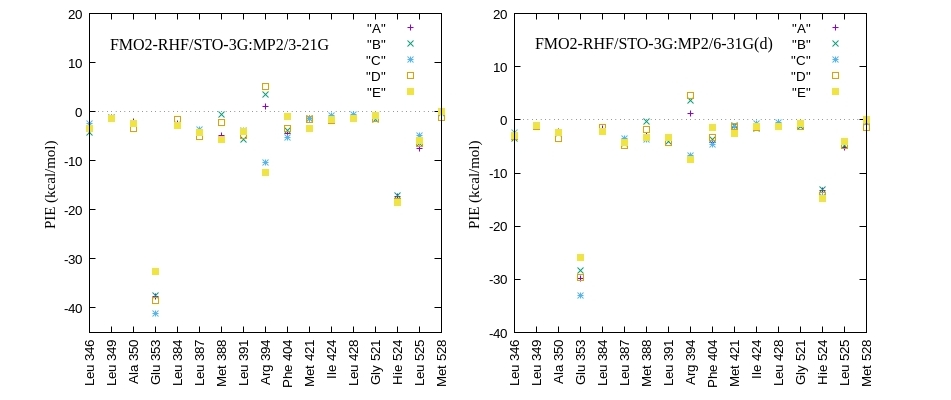}
\caption{PIEs (kcal/mol) between 17$\beta$-estradiol and each amino acid residue fragment of the ER binding site for each of the representative structures. Glu 353, Arg 394 and Hie 524 are charged/polar residues; the remaining residues are hydrophobic.}
\label{figure5}
\end{figure*} 
 
During the simulation a number of water molecules were trapped at the binding site. Table V shows the interaction energy calculated at the FMO2-RHF/STO-3G level of theory between E2 and each water fragment in the ER binding site. The number of water molecules varies according to the representative structure and, at this level of calculation, the interactions can be either attractive or repulsive. It was observed that only in the representative structure of the most populated cluster (A) a water molecule makes a bridge between E2 and Leu 387 through the hydrogen bonds ($\Delta E_{int}$ = -4.05 kcal/mol). The negative value in the interaction energy of the other structures (C, E) corresponds to an interaction by hydrogen bond between the 17$\beta$-OH group (D ring) of E2 and a water molecule.

One shortcoming of this study is that we do not consider the binding site as a dynamic entity, subject to change. These, are intrinsic to the dynamics of the system itself. The binding site is defined by the distance between E2 and the corresponding amino acid residue, and this distance is modified in the course of the simulation. Therefore, some residues will cease to belong to the binding site, while others, which initially did not belong, will become part of the binding site (Table VI). The above is also relevant for water molecules. We hope to address this inconsistency in a future study.

\begin{table}[!htbp]
  \begin{threeparttable}
\caption{FMO2-RHF/STO-3G interaction energy$^{(a)}$ ($\Delta E_{int}$).}
\begin{tabular}{cccccc}
  \toprule[1pt]
  RS$^{(b)}$ & \ \ \ \ A & \ \ B & \ \ C & \ \ D & \ \ E \\
  \midrule[1pt]
  Water fragment & \ \ \ \ 4 & \ \ 1 & \ \ 3 & \ \ 2 & \ \ 3 \\
  $\Delta E_{int}$ & \ \ \ \ -4.05 & \ \ 0.87 & \ \ -4.01 & \ \ 1.38 & \ \ -4.80 \\
  \bottomrule[1pt]
\end{tabular}
\begin{tablenotes}
      \small
      \item $^{(a)}$Sum of all PIEs between E2 and each water fragment in the ER binding site. $^{(b)}$Representative structure. All Energies in kcal/mol.
\end{tablenotes}
\label{Table V}
  \end{threeparttable}
\end{table}

\begin{table*}[!htbp]
  \begin{threeparttable}
\caption{Effect of the simulation on the amino acid residue$^{(a)}$ composition of the ER binding site. The following original residues remained at the binding site: Leu 346 $\rightarrow$ Leu 525. The binding site consists of all residues that have at least one atom within 3.5 \r{A} from any E2 atom in E2-ER-w model system. ($\Delta E_{int}$).}
\begin{tabular}{cccccc}
  \toprule[1pt]
  \multirow{1}{*}{RS} & \ \ \ \ \ A & \ \ \ \ \ B & \ \ \ \ \ C & \ \ \ \ \ D & \ \ \ \ \ E \\
  \midrule[1pt]
  \multirow{1}{*}{ } & \ \ \ \ \  & \ \ \ \ \ \ \ \ Met 528$^{(b)}$ & \ \ \ \ \ Met 528 & \ \ \ \ \ Met 528 & \\
  \multirow{1}{*}{ } & \ \ \ \ \ Met 343 & \ \ \ \ \ Met 343 & \ \ \ \ \ Met 343 & \ \ \ \ \ Met 343 & \\
  \multirow{1}{*}{Aa residue} & \ \ \ \ \  & \ \ \ \ \  & \ \ \ \ \  & \ \ \ \ \  & \ \ \ \ \ Leu 327 \\
  \multirow{1}{*}{ } & \ \ \ \ \ Phe 425 &  &  &  &  \\
  \multirow{1}{*}{ } & \ \ \ \ \ \underline{Thr 347} & \ \ \ \ \  & \ \ \ \ \ \underline{Thr 347} &  & \\
  \multirow{1}{*}{ } & \ \ \ \ \ Trp 383 &  &  &  & \\
  \bottomrule[1pt]
\end{tabular}
\begin{tablenotes}
      \small
      \item $^{(a)}$Charged/Polar residues are indicated by underlined characters. The remaining residues are hydrophobic. $^{(b)}$Original amino acid residue.
\end{tablenotes}
\label{Table VI}
  \end{threeparttable}
\end{table*}

We have shown that if the structure representing the free receptor (ER-w) differs significantly from that of the bound receptor (E2-ER-w), the binding energy calculations are not reliable. However, if we obtain the free receptor from the respective bound receptor, the values of the binding energy are less dispersed and seemingly reliable. In this particular case, the backbone RMSD is very small between both representative structures; that is, the conformations are very similar. The justification for the second calculation scheme is somewhat supported by the conformational selection model, which postulates that the native state of a protein does not exist as a single, rigid conformation but rather as a ensemble of conformers that coexist in equilibrium with different population distributions, and that the ligand can bind selectively to the most suitable conformer. In other words, the free protein can sample with a certain probability the same conformation as that of the bound protein \cite{62}. If we accept the above hypothesis, it is permissible to derive the free receptor from the bound receptor. Finally, we must mention that although it is true that the representative structures obtained from the independent MD simulations of both systems differ significantly from each other and, therefore, contradict one of the premises of the conformational selection, the reason for this possibly lies in the fact that the extension of the sampled space is insufficient for the free system. Increasing conformational sampling of the free receptor will likely allow to obtain structures more similar to those of the bound receptor.

\section{Conclusions}

The molecular interactions between 17$\beta$-estradiol and ER were calculated, and from these the binding energy was obtained. Two different schemes were used for the calculation of the binding energy. In the first scheme, the free receptor used in the calculations is obtained from MD simulations, and the results are not reliable as the energy values are highly dispersed. This failure is a consequence of the lack of structural similarity between the representative structures of the free and the bound receptor. The second scheme, which produces reliable results, uses the structure of the free receptor obtained directly from the bound receptor by removal of E2. The above procedure is consistent with the conformational selection model. We think that by increasing the conformational sampling of the free receptor it is possible to obtain structures similar to those of the bound receptor; indeed, if this were the case, we could provide reasonable theoretical evidence in favor of the conformational hypothesis.

In general, attractive dispersion interactions were observed between E2 and all surrounding hydrophobic residues. These interactions play an important role in stabilizing E2 at the binding site. Water molecules were found at the binding site of all representative structures; in one of them (A), a water molecule makes a bridge between E2 and Leu 387 through the hydrogen bonds. Strong electrostatic interactions were observed between the E2 and the following charged/polarized residues: Glu 353, His 524 and Arg 394. These residues tend to be located at the ends of E2, close to the OH groups of the A and D rings. In particular, the behavior of Arg 394 is quite unusual since depending on its conformation or position in the active site, the interaction with E2 can be attractive or repulsive.

The FMO2-RHF/STO-3G:MP2/6-31G(d) weighted binding energy was of -67.2 kcal/mol. This result clearly suggests that 17$\beta$-estradiol and ER LBD tend to form an enzyme-substrate complex and therefore validates our calculation model based on interactions. We hope that the model developed in this study can be useful for identifying and assessing the health risk of potential EDCs.

\section*{Acknowledgements}

I wish to express my gratitude to each and every person who has contributed to the development and maintenance of free and open source software. I thank Dr. Robin Mesnage for English assistance and constructive comments on the manuscript.

\bibliographystyle{plain}
\bibliography{manuscrito}

\end{document}